\documentclass[showpacs,preprintnumbers,
amsmath,amssymb,nofootinbib]{revtex4}
\usepackage{graphicx}
\usepackage{dcolumn}
\usepackage{bm}

\begin{document}


\title{Reply by N. Dinh Dang to the Comment by A. Rabhi}

\author{Nguyen Dinh Dang}
\affiliation{%
RI-beam factory project office, RIKEN, 2-1 Hirosawa, Wako, 351-0198 
 Saitama, Japan}%
\date{\today}
\begin{abstract}
It is pointed out that the ``exact'' solutions given in the recent Comment 
by A. Rabhi on the paper \lbrack N. Dinh Dang ``Energies of the ground 
state and first excited $0^{+}$ state in an exactly solvable pairing 
model'', Eur. Phys. Jour. A {\bf 16} (2003) 181\rbrack
~~do not correspond to the standard exact 
solution of the well-known two-level pairing model. Other issues 
raised in the Comment reiterate the discussions already published 
in the original paper by N. Dinh Dang. 
\end{abstract}
\maketitle
In Ref. \cite{Dang} several approximations, namely the 
BCS approximation, Lipkin - Nogami method, random-phase 
approximation (RPA), quasiparticle RPA (QRPA), the renormalized RPA (RRPA), 
and renormalized QRPA (RQRPA), are tested by calculating 
the ground-state energy and the energy of the 
first excited $0^{+}$ state  using the well-known exactly solvable model with 
two symmetric levels interacting via a pairing force. 
The author of the recent Comment \cite{Comment} discusses three points 
of Ref. \cite{Dang}, namely (A) the boson mapping in Sec. 4.1.1, 
(B) the approximations leading to the RPA matrices (58) -- (61) and 
those obtained in Ref. \cite{Hagino}, and (C) the approximation (75).

(A) The author of \cite{Comment} claimed that
the one-boson energy $\omega_{\rm RPA}^{(b)}$ 
given by Eq. (53) of \cite{Dang} 
is erroneous. As the proof he introduces two ``exact'' solutions for 
the $0^{+}$ energy in Fig. 1 of \cite{Comment}. However, these ``exact'' 
solutions are completely different from the standard exact solution of 
the well-known two-level model under consideration, which has been 
obtained using the SU(2) algebra in many papers [See Sec. 2 and Fig. 3 
of Ref. \cite{Dang}, or Refs. \cite{Hogaasen,Samba}, e.g.]. In particular, as 
has been mentioned at the end of Sec. 4.1.1 of \cite{Dang}, the 
solution given by Eq. (53) of \cite{Dang} 
is exactly the same as that 
obtained for the first time by Hogaasen-Fledman in Ref. \cite{Hogaasen} using
the space-variable technique. The author of the Comment \cite{Comment}, however, 
fails to reproduce the boson and exact solutions by 
H\"{o}gaasen-Feldman, saying that he does not understand it
\footnote{The derivation in Eqs. (2) - (10) in the Comment \cite{Comment} was 
actually sent by Dinh Dang to Rabhi after the latter failed to reproduce the 
result of Eq. (53) of \cite{Dang} as well as that obtained by Hogaasen 
in \cite{Hogaasen}}.

(B) Sec. 4.1.2 just discusses two approximations (62) and (63). The 
former is based on the exact commutation relations in Eq. (10) and 
leads to the matrices (58) -- (61). The latter leads to those
in Ref. \cite{Hagino}. The difference is the factor of 2 in the 
denominator of the second term at the right-hand side of 
Eq. (58) in \cite{Dang} 
and the corresponding term in Eq. (38) of \cite{Hagino}. 
The omission of $q$-term was considered in \cite{Dang} also as a 
possible approximation, which leads to the appearance of the 
spurious mode as has been pointed out in \cite{Dang} and repeated by 
the author of \cite{Comment}. It is worthwhile to study this 
approximation since there have been numerical calculations within the RPA 
neglecting the so-called scattering terms, which have the same origin 
as that of the $q$-term considered here (See the discussion in (b) on 
page 185 of \cite{Dang}). In such calculations the parameters of the 
effective interaction are usually readjusted in such a way that that
the energy of the spurious mode is zero to compensate for such 
effect [See, e.g. Ref. \cite{Soloviev}]. 
To my knowledge, there is no exact way to take into account the 
scattering term within 
the QRPA so far. Approximations to take into account the scattering term 
lead to the extended QRPA~\cite{ERPA} 
or modified QRPA~\cite{MRPA}.

(C) The approximation (75) in \cite{Dang} has been introduced so that 
one can compare the exact solution, the 
phonon solution obtained within the fermion 
formalism in Sec. 4.1.2 
with that obtained within the one-boson mapping in Sec. 4.1.1.
The conclusion is that the exact solution can be approximately 
considered as a mixture of superfluid and normals fluid states as
developed in Sec. 4.3. In this sense the approximation
(75) mixes $N \pm 2$ states.

The RPA operator for the additional mode in the present two-level 
model is~\cite{Ring}
\begin{equation}
    Q_{\rm a}^{\dagger}=X_{2}A_{2}^{\dagger}-Y_{1}A_{1}^{\dagger}~.
    \label{Qa}
    \end{equation}
The one for the removal mode is
\begin{equation}
    Q_{\rm r}^{\dagger}=X_{1}A_{1}^{\dagger}-Y_{2}A_{2}^{\dagger}~.
    \label{Qr}
    \end{equation}
It is clear that 
neither $Q_{\rm a}^{\dagger}$ nor $Q_{\rm r}^{\dagger}$ conserves the 
particle number on each level because, according to the exact 
commutation relations in Eq. (5) of \cite{Dang}, 
they do not commute with $N_{i}$ ($i=$1,2), 
except for $X_{i}=$ 0 or $Y_{i}=$ 0. So doesn't the operator
$Q^{\dagger}=Q_{\rm a}^{\dagger}+Q_{\rm r}^{\dagger}$ in Eq. (75) 
except for
$X_{i}=Y_{i}=$ 0.
However, in the boson formalism, based on the mapping (49) and (50) 
of \cite{Dang} with ${\bf b}_{1}={\bf b}_{2}={\bf b}$ as has been 
discussed in the Comment \cite{Comment}, one 
obtains the boson images of these commutation relations as follows
\begin{subequations}
 \label{commu}   
\begin{equation}
    [N_{1},Q_{\rm a}^{\dagger}]_{\rm b}=[2\Omega-2{\bf 
    b}^{\dagger}{\bf b},X_{2}{\bf b}^{\dagger}-Y_{1}{\bf 
    b}]=-2X_{2}{\bf b}^{\dagger}-2Y_{1}{\bf b}~,
    \label{commu1}
    \end{equation}
\begin{equation}
    [N_{2},Q_{\rm a}^{\dagger}]_{\rm b}=[2{\bf 
    b}^{\dagger}{\bf b},X_{2}{\bf b}^{\dagger}-Y_{1}{\bf 
    b}]=2X_{2}{\bf b}^{\dagger}+2Y_{1}{\bf b}~,
    \label{commu2}
\end{equation}    
\begin{equation}
    [N_{1},Q_{\rm r}^{\dagger}]_{\rm b}=[2\Omega-2{\bf 
    b}^{\dagger}{\bf b},X_{1}{\bf b}-Y_{2}{\bf 
    b}^{\dagger}]=2X_{1}{\bf b}+2Y_{2}{\bf b}^{\dagger}~,
    \label{commu3}
\end{equation}    
\begin{equation}
    [N_{2},Q_{\rm r}^{\dagger}]_{\rm b}=[2{\bf 
    b}^{\dagger}{\bf b},X_{1}{\bf b}-Y_{2}{\bf 
    b}^{\dagger}]=-2X_{1}{\bf b}-2Y_{2}{\bf b}^{\dagger}~,
    \label{commu4}
\end{equation}    
\end{subequations}
Summing up Eqs. (\ref{commu1}) and (\ref{commu2}), one finds
\begin{equation}
    [N_{1}+N_{2},Q_{\rm a}^{\dagger}]_{\rm b}=
    [N,Q_{\rm a}^{\dagger}]_{\rm b}=0~.
    \label{1+2}
    \end{equation}
Summing up Eqs. (\ref{commu3}) and (\ref{commu4}), one finds
\begin{equation}
    [N_{1}+N_{2},Q_{\rm r}^{\dagger}]_{\rm b}=
    [N,Q_{\rm r}^{\dagger}]_{\rm b}=0~.
    \label{3+4}
    \end{equation}
Hence
\begin{equation}
    [N,Q^{\dagger}]=[N,Q_{\rm a}^{\dagger}+Q_{\rm r}]=0~.
    \label{a+r}
    \end{equation}
These results show that, in the boson formalism, although the 
additional and removal operators do not conserve the particle number 
$N_{i}$ on each level, they do conserve the total particle number
$N=N_{1}+N_{2}$. So does the operator $Q^{\dagger}=Q_{\rm a}^{\dagger}+
Q_{\rm r}^{\dagger}$ because of Eq. (\ref{a+r}).

In conclusion, the present Reply clarified several issues raised in 
the recent Comment \cite{Comment} on the paper \cite{Dang} regarding 
the RPA and QRPA for an exactly solvable pairing model.

\end{document}